\documentclass[pra,aps,twocolumn,superscriptaddress,showpacs]{revtex4}

\usepackage[cp1250]{inputenc}
\usepackage{graphics}

\begin{document}

\title{Transverse coherence of photon pairs generated in spontaneous parametric down-conversion}

\author{Martin Hamar}

\author{Jan Pe\v{r}ina Jr.}

\author{Ond\v{r}ej Haderka}

\author{V\'{a}clav Mich\'{a}lek}

\affiliation{Joint Laboratory of Optics of Palack\'{y} University
and Institute of Physics of Academy of Sciences of the Czech
Republic, 17. listopadu 50a, 772 07 Olomouc, Czech Republic}

\begin{abstract}
Coherence properties of the down-converted beams generated in
spontaneous parametric down-conversion are investigated in detail
using an iCCD camera. Experimental results are compared with those
from a theoretical model developed for pulsed pumping with a
Gaussian transverse profile. The results allow to tailor the shape
of correlation area of the signal and idler photons using
pump-field and crystal parameters. As an example, splitting of a
correlation area caused by a two-peak pump-field spectrum is
experimentally studied.

\end{abstract}

%\pacs{03.67.Lx, 42.50.-p}

\maketitle

\section{Introduction}

Light emitted from spontaneous parametric down-conversion in a
nonlinear crystal is composed of photon pairs. Two photons
comprising a photon pair are called a signal and an idler photon
for historic reasons. The first theoretical investigation of this
process has been done in year 1968 \cite{Giallorenziho1968}.
Already this study has revealed that frequencies and emission
directions of two photons in a pair are fully determined by the
laws of energy and momentum conservations. For this reason, there
occurs a strong correlation (entanglement) between properties of
the signal and idler photons. In an ideal case of infinitely long
and wide nonlinear crystal and monochromatic plane-wave pumping, a
plane-wave signal photon at frequency $ \omega_s $ belongs just to
one plane-wave idler photon at frequency $ \omega_i $ that is
determined by the conservation of energy. Emission angles of these
photons are given by the momentum conservation that forms
phase-matching conditions. Possible signal (and similarly idler)
emission directions lie on a cone which axis coincides with the
pump-beam direction of propagation.

However, real experimental conditions have enforced the
consideration of crystals with finite dimensions
\cite{Hong1985,Wang1991}, pump beams with nonzero divergence
\cite{Grayson1994,Steuernagel1998} as well as pulsed pumping
\cite{Keller1997,DiGiuseppe1997,Grice1997,PerinaJr1999}. During
this investigation, the approximation based on a multidimensional
Gaussian spectral two-photon amplitude has been found
extraordinarily useful \cite{Joobeur1994,Joobeur1996}. The
developed models have revealed that spatial characteristics of a
pump beam are transferred to certain extent to these of a photon
pair generated in a nonlinear crystal, especially in case of short
crystals
\cite{Nasr2002,Walborn2004,Monken1998,Molina-Terriza2005}. These
models have also been useful in quantifying real effects in
applied experimental setups utilizing photon pairs
\cite{Shih2003,Law2004}. They have also been recently extended to
photonic \cite{Centini2005,PerinaJr2006} and wave-guiding
\cite{Ding1995,DeRossi2002,Booth2002,Walton2003,Walton2004,PerinaJr2008}
structures. Also effects at nonlinear boundaries have been taken
into account \cite{PerinaJr2009,PerinaJr2009a}.

In this paper, we continue the previous investigations of spatial
photon-pair properties
\cite{Saleh1998,Howell2004,DAngelo2004,Brambilla2004} by
experimental study of transverse profiles of the down-converted
beams as well as correlation areas of the signal and idler photons
using an iCCD camera \cite{Jost1998,Haderka2005,Haderka2005a}.
Special attention is paid to the role of pump-beam parameters.
Experimental results are compared with a theoretical model that
considers Gaussian spectrum and elliptical pump-beam profile. We
note that also sensitive CCD cameras have been found useful in
investigations of spatial properties of more intense twin beams
\cite{Jiang2003,Jedrkiewicz2004,Jedrkiewicz2006}.

The paper is organized as follows. A theoretical model is
presented in Sec.~II. Sec.~III brings theoretical analysis of
parameters of a correlation area as well as spectral properties of
the down-converted fields. An experimental method based on the use
of an iCCD camera is discussed in detail in Sec.~IV. The
experimentally observed dependence of parameters of the
correlation area on pump-beam characteristics and crystal length
is reported in Sec.~V. Sec.~VI is devoted to splitting of the
correlation area and its experimental observation. Conclusions are
drawn in Sec.~VII.

\section{Theory}

The process of spontaneous parametric down-conversion is described
by the following interaction Hamiltonian $ \hat{H}_{\rm int} $
\cite{Hong1985,Saleh1991,Shih2003}:
\begin{eqnarray}      % 1
 \hat{H}_{\rm int}(t) &=& \varepsilon_0 \int\limits_V d{\bf r}
 \chi^{(2)} : {\bf E}_{p}^{(+)}({\bf r},t) \hat{\bf E}_{s}^{(-)} ({\bf r},t)
 \hat{\bf E}_{i}^{(-)}({\bf r},t) \nonumber \\
 & & \mbox{} + {\rm H.c.} ,
\label{1}
\end{eqnarray}
where $ {\bf E}_{p}^{(+)} $ is the positive-frequency part of the
pump electric-field amplitude, whereas $ {\bf E}_{s}^{(-)} $ ($
{\bf E}_{i}^{(-)} $) stands for the negative-frequency part of the
signal (idler) electric-field amplitude operator. Symbol $
\chi^{(2)} $ means the second-order susceptibility tensor and $ :
$ is shorthand for tensor reduction with respect to its three
indices. Susceptibility of vacuum is denoted as $ \varepsilon_0 $,
interaction volume as $ V $ and $ {\rm H.c.} $ substitutes the
Hermitian-conjugated term.

We further consider parametric down-conversion in a LiIO$ {}_3 $
crystal with an optical axis perpendicular to the $ z $ axis of
fields' propagation direction and type-I interaction. The pump
field is assumed to be polarized vertically (it propagates as an
extraordinary wave) whereas the signal and idler fields are
polarized horizontally (they propagate as ordinary waves). In this
specific configuration, scalar optical fields are sufficient for
the description. The interacting optical fields can then be
decomposed into monochromatic plane waves with frequencies $
\omega_a $ and wave vectors $ {\bf k}_a $:
\begin{eqnarray}     % 2
 E_a^{(+)}({\bf r},t) &=& \int d{\bf k}_a E_a^{(+)}({\bf k}_a)
  \exp(i{\bf k}_a{\bf r}-i\omega_a t) + {\rm H.c.}; \nonumber \\
 & &   \hspace{3cm}  a=p,s,i.
\label{2}
\end{eqnarray}
The signal and idler fields at a single-photon level have to be
described quantally and so their spectral amplitudes $
\hat{E}_a^{(+)}({\bf k}_a) $ can be expressed as $
\hat{E}_a^{(+)}({\bf k}_a) = i
\sqrt{\hbar\omega_a}/\sqrt{2\varepsilon_0 c {\cal A}
n_a(\omega_a)} \hat{a}_a({\bf k}_a) $ using annihilation operators
$ \hat{a}_a({\bf k}_a) $ that remove one photon from a plane-wave
mode $ {\bf k}_a $ in field $ a $. Symbol $ \hbar $ stands for the
reduced Planck constant, $ c $ is speed of light in vacuum, $
{\cal A} $ transverse area of a beam, and $ n_a $ means index of
refraction in field $ a $.

Under these conditions, the interaction Hamiltonian $ \hat{H}_{\rm
int} $ in Eq.~(\ref{1}) takes the form \cite{Joobeur1994}:
\begin{eqnarray}    % 3
 \hat{H}_{\rm int}(t) &=& A_n(\omega_s^0,\omega_i^0)
  \int d\mathbf{k}_s \int d\mathbf{k}_i \int d\mathbf{k}_p
  E_{p}^{(+)}(\mathbf{k}_p ) \nonumber  \\
 & & \mbox{} \times \exp\left\{i\left[ \omega(\mathbf{k}_p) - \omega(\mathbf{k}_s) -
  \omega(\mathbf{k}_i) \right] t \right\} \nonumber  \\
 & & \mbox{} \times \int\limits_V d\mathbf{r}
   \exp\left[-i\left(\mathbf{k}_p - \mathbf{k}_s  -
   \mathbf{k}_i \right)\mathbf{r}\right]  \nonumber \\
 & & \mbox{} \times \hat{a}^{\dagger}_s(\mathbf{k}_s)
  \hat{a}^{\dagger}_i(\mathbf{k}_i) + {\rm H.c.}
\label{3}
\end{eqnarray}
We have assumed in deriving Eq.~(\ref{3}) that the function $ A_n
(\omega_s,\omega_i) = - \hbar\sqrt{\omega_s\omega_i} \chi^{(2)} /
(2c{\cal A} \sqrt{n_s(\omega_s) n_i(\omega_i)} $ is a slowly
varying function of frequencies $ \omega_s $ and $ \omega_i $ and
can be approximated by its value taken at the central frequencies
$ \omega_s^0 $ and $ \omega_i^0 $.

A quantum state $ | \Psi  \rangle $ of a generated photon pair can
be obtained after solving the Schr\"{o}dinger equation up to the
first power of the interaction constant that results in the
formula:
\begin{equation}    % 4
 | \Psi  \rangle  =  - \frac{i}{\hbar}
  \int_{-\infty}^{\infty} dt \hat{H}_{\rm int} (t) | {\rm vac}
  \rangle;
\label{4}
\end{equation}
$ |{\rm vac} \rangle $ means the vacuum state. Substitution of the
interaction Hamiltonian $ \hat{H}_{\rm int} $ from Eq.~(\ref{3})
into Eq.~(\ref{4}) provides the following form for the quantum
state $ |\Psi\rangle $
\cite{Grayson1994,Shih2003,Ou1989,Mandel1995}:
\begin{equation}   % 5
 | \Psi  \rangle  = \int d \mathbf{k}_{s}
  \int d \mathbf{k}_{i} S(\mathbf{k}_{s},\mathbf{k}_{i})
  \hat{a}_s^\dagger({\bf k}_s) \hat{a}_i^\dagger({\bf k}_i)
  |{\rm vac}\rangle ,
\label{5}
\end{equation}
where the newly introduced two-photon amplitude $ S $ takes the
form:
\begin{eqnarray}  % 6
 S(\mathbf{k}_{s},\mathbf{k}_{i}) &=& A'_n \int d \mathbf{k}_{p}
  E_{p}^{( + )}(\mathbf{k}_{p}) \delta(\omega_p  - \omega_s  - \omega_i)
  \nonumber \\
 & & \times \int_V d \mathbf{r} \exp\left[-i(
  \mathbf{k}_{p}  - \mathbf{k}_{s}  - \mathbf{k}_{i})
  \mathbf{r} \right]
\label{6}
\end{eqnarray}
and $ A'_n = -2\pi i / \hbar A_n(\omega_s^0,\omega_i^0) $. We note
that squared modulus $ |S({\bf k}_s,{\bf k}_i)|^2 $ of the
two-photon amplitude gives us the probability density of
simultaneous generation of a signal photon with wave vector $ {\bf
k}_s $ and its twin with wave vector $ {\bf k}_i $.

Spectral resolution is usually not found in experiments with
photon pairs and then the photon-pair coincidence-count rate is
linearly proportional to the fourth-order correlation function $
G_{s,i} $ defined as:
\begin{eqnarray}   % 7
 G_{s,i}(\xi_s,\delta_s,\xi_i,\delta_i) &=& \frac{
  \sin(\xi_s)\sin(\xi_i) }{c^6} \int d\omega_s
  \omega_s^2 \nonumber \\
 & & \hspace{-3cm} \times  \int d\omega_i \omega_i^2
  |h(\omega_s) h(\omega_i)|^2 |S(\xi_s,\delta_s,\omega_s,
  \xi_i,\delta_i,\omega_i)|^2;
\label{7}
\end{eqnarray}
$ S(\xi_s,\delta_s,\omega_s,\xi_i,\delta_i,\omega_i) \equiv S({\bf
k}_s,{\bf k}_i) $. The propagation direction of a photon is
parameterized by radial emission angles $ \xi_a $ (determining
declination from the $ z $ axis) and azimuthal emission angles $
\delta_a $ (describing rotation around the $ z $ axis starting
from the $ x $ axis); $ a=s,i $ (see also Fig.~\ref{fig8}).
Functions $ h_s $ and $ h_i $ introduced in Eq.~(\ref{7}) describe
amplitude spectral and/or geometrical filtering of photons in
front of detectors.

More detailed information is contained in intensity spectrum $ S_s
$ of a signal field assuming photon pairs emitted into the fixed
signal- and idler-photon directions given by angles $ \xi_s $, $
\delta_s $, $ \xi_i $, and $ \delta_i $:
\begin{eqnarray}   % 8
 S_s(\omega_s;\xi_s,\delta_s,\xi_i,\delta_i) &=& \frac{
 \sin(\xi_s)\sin(\xi_i) \omega_s^2 |h(\omega_s)|^2}{c^6} \nonumber \\
 & & \hspace{-2.5cm} \times  \int d\omega_i \omega_i^2
  |h(\omega_i)|^2 |S(\xi_s,\delta_s,\omega_s,
  \xi_i,\delta_i,\omega_i)|^2.
\label{8}
\end{eqnarray}
If the signal-photon emission direction described by angles $
\xi_s $ and $ \delta_s $ is not resolved, an integrated
signal-field emission spectrum $ S_s^{\rm int} $ is observed:
\begin{equation}  % 9
 S_s^{\rm int}(\omega_s;\xi_i,\delta_i) = \int_{-\pi/2}^{\pi/2}
 d\xi_s \int_{-\pi}^{\pi} d\delta_s
 S_s(\omega_s;\xi_s,\delta_s,\xi_i,\delta_i).
\label{9}
\end{equation}
Similar formulas as given in Eqs.~(\ref{8}) and (\ref{9}) can be
derived also for the idler field.

On the other hand excluding resolution in emission directions,
spectral correlations between the signal and idler fields are
characterized by a two-photon spectral amplitude $ \Phi_{s,i} $
which squared modulus is defined as:
\begin{eqnarray}   % 10
 |\Phi_{s,i}(\omega_s,\omega_i)|^2 &=& \frac{\omega_s^2
  \omega_i^2}{c^6} \int d\delta_s \int d\xi_s
  \int d\delta_i \int d\xi_i  \nonumber
  \\
 & & \hspace{-1cm} \sin(\xi_s) \sin(\xi_i) |h(\omega_s)
  h(\omega_i)|^2 \nonumber \\
 & & \hspace{-1cm} \mbox{} \times |S(\xi_s,\delta_s,\omega_s,\xi_i,\delta_i,\omega_i)|^2.
\label{10}
\end{eqnarray}

We further consider a Gaussian pump beam with the electric-field
amplitude $ E_{p}^{(+)} $ in the from:
\begin{eqnarray}     % 11
 E_p^{(+)}({\bf r},t) &=& \int d\omega_p A_p(\omega_p)\exp(i{\bf k}_{pz}z -i\omega_p t)
  \nonumber \\
  & & \hspace{-10mm} \times \frac{1}{W_{px}(z)} \exp\left[ -\frac{x^2}{W_{px}^2(z)} \right]
   \exp\left[ -ik_p\frac{x^2}{2R_{px}^2(z)} \right] \nonumber \\
  & & \hspace{-10mm} \times \frac{1}{W_{py}(z)} \exp\left[ -\frac{y^2}{W_{py}^2(z)} \right]
   \exp\left[ -ik_p\frac{y^2}{2R_{py}^2(z)} \right] \nonumber \\
  & & \hspace{-10mm} \times \exp[i\zeta_p(z)];
\label{11}
\end{eqnarray}
$ k_p = |{\bf k}_p| $. The functions $ W_{pa} $, $ R_{pa} $, and $
\zeta_p $ are defined as:
\begin{eqnarray}   % 12-14
 W_{pa}(z) &=& W_{pa}^0 \sqrt{ 1+ \frac{z^2}{(z_{pa}^0)^2} },
  \hspace{2mm} W_{pa}^0 = \sqrt{\frac{2z_{pa}^0}{k_p}}, \\
 R_{pa}(z) &=& z \left[ 1+ \frac{(z_{pa}^0)^2}{z^2} \right] ,
  \hspace{10mm} a=x,y ,\\
 \zeta_p(z) &=& [\arctan(z/z_{px}^0) + \arctan(z/z_{py}^0)]/2.
\end{eqnarray}
The function $ A_p $ introduced in Eq.~(\ref{11}) gives the
pump-field amplitude temporal spectrum. Constants $ z_{px}^0 $ and
$ z_{py}^0 $ describe positions of waists with radii $ W_{px}^0 $
and $ W_{py}^0 $ in the $ x $ and $ y $ directions, respectively.
Function $ W_{px}(z) $ [$ W_{py}(z) $] gives a radius of the beam
in the $ x $ [$ y $] direction and with wavefront curvature $
R_{px}(z) $ [$ R_{py}(z) $] at position $ z $.

We assume that the nonlinear crystal is sufficiently short so that
changes of the pump-field amplitude $ E_p^{(+)} $ in the
transverse plane along the $ z $ axis can be neglected. In this
case, the pump-field amplitude $ E_p^{(+)} $ can be characterized
both by its temporal spectrum $ A_p(\omega_p) $ and spatial
spectrum $ F_{p}(k_{px},k_{py}) $ in the transverse plane:
\begin{eqnarray}   % 15
 E_p^{(+)}({\bf r},t) &=& \int d\omega_p A_p(\omega_p)\int d{\bf k}_{px}
  \int d{\bf k}_{py}  \nonumber \\
 & &  \hspace{-15mm}  F_p({\bf k}_{px},{\bf k}_{py}) \exp(i{\bf k}_{px}x)
  \exp(i{\bf k}_{py}y) \nonumber \\
 & & \hspace{-15mm} \times  \exp(i{\bf k}_{pz}z)  \exp(-i\omega_p t);
\label{15}
\end{eqnarray}
The spatial spectrum $ F_p $ corresponding to the Gaussian beam
written in Eq.~(\ref{11}) and propagating along the $ z $ axis can
be expressed as:
\begin{eqnarray}  % 16
 F_p({\bf k}_{px},{\bf k}_{py}) &=& \frac{1}{W_{px}(z_0) W_{py}(z_0)}
  \frac{2}{ \bar{W}_{px} \bar{W}_{py} } \nonumber \\
 & & \hspace{-20mm} \times \exp\left[ -\frac{{\bf k}_{px}^2}{\bar{W}_{px}^2}
 \right] \exp\left[ -\frac{{\bf k}_{py}^2}{\bar{W}_{py}^2}
 \right] \exp[i\zeta_p(z_0)],
\label{16}
\end{eqnarray}
where the position $ z_0 $ lies inside the crystal. Complex
spectral half-widths $ \bar{W}_{px} $ and $ \bar{W}_{py} $ of the
spatial spectrum in the transverse plane are given as follows:
\begin{equation}   % 17
 \bar{W}_{pa} = 2\sqrt{ \frac{1}{W_{pa}^2(z_0)} +
 \frac{ik_p}{2R_{pa}^2(z_0)} }, \hspace{5mm} a=x,y .
\end{equation}

In the following we consider a Gaussian chirped pump pulse which
temporal amplitude spectrum $ A_p $ can be expressed in the form:
\begin{equation}    % 18
 A_p(\omega_p) = \xi_p \frac{\tau_p}{\sqrt{2(1+ia_p)}}
  \exp\left[ -\frac{\tau_p^2}{4(1+ia_p)} \omega_p^2 \right].
\label{18}
\end{equation}
In Eq.~(\ref{18}), $ \tau_p $ denotes pump-pulse duration, $ a_p $
stands for a chirp parameter, and $ \xi_p $ is the pump-field
amplitude. We note that the amplitude width $ \Delta\omega_p $
(given as full width at $ 1/e $ of the maximum) of the pulse
written in Eq.~(\ref{18}) equals $ 4\sqrt(1+a_p^2)/\tau_p $.

Considering the pump-field amplitude $ E_p^{(+)} $ as given in
Eq.~(\ref{15}) the two-photon amplitude $ S $ defined in
Eq.~(\ref{6}) can be recast into the form:
\begin{eqnarray}    % 19
 S(\xi_s,\delta_s,\omega_s, \xi_i,\delta_i,\omega_i) &=&
  c A_p(\omega_s+\omega_i) \nonumber \\
  & & \hspace{-32mm} \times F_p({\bf k}_{sx}+{\bf k}_{ix},{\bf k}_{sy}+{\bf k}_{iy})
   \nonumber \\
  & & \hspace{-32mm} \times L_z {\rm sinc} \left\{
   \frac{[{\bf k}_{pz}(\omega_s+\omega_i)-{\bf k}_{sz}(\omega_s)-{\bf k}_{iz}(\omega_i)]L_z}{2}
   \right\} \nonumber \\
  & & \hspace{-32mm} \times \exp \left\{ -i
   \frac{[{\bf k}_{pz}(\omega_s+\omega_i)-{\bf k}_{sz}(\omega_s)-{\bf k}_{iz}(\omega_i)]L_z}{2}
   \right\} ; \nonumber \\
   & &
\label{19}
\end{eqnarray}
$ {\rm sinc}(x) = \sin(x)/x $. In deriving Eq.~(\ref{19}), we have
assumed that the crystal extents from $ z = -L_z $ to $ z = 0 $, $
L_z $ being the crystal length. The transverse profile of crystal
is also assumed to be sufficiently wide.

\section{Correlation area, spectral properties}

Correlation area is defined by the profile of probability density
of detecting a signal photon in the direction described by angles
($\xi_s,\delta_s $) provided that its idler twin has been detected
in a fixed direction given by angles ($\xi_i,\delta_i $). In
coherence theory, this probability is given by the fourth-order
correlation function $ G_{s,i} $ defined in Eq.~(\ref{7}). Because
the correlation function $ G_{s,i} $ is usually a smooth function
of its arguments, it can be conveniently parameterized using
angular widths (given as full-widths at $ 1/e $ of maximum) in the
radial ($ \Delta\xi_s $) and azimuthal ($ \Delta\delta_s $)
directions. In general, parameters of the correlation area depend
on properties of crystal material as well as crystal length,
pump-field spectral bandwidth, and transverse pump-beam profile.
The last two parameters allow to tailor characteristics of the
correlation area in wide ranges.

In the theoretical analysis of Sec.~III, we use radial ($ \xi $)
and azimuthal ($ \delta $) angles inside a nonlinear crystal. The
reason is that we want to exclude the effect of mixing in spatial
and frequency domains at the output plane of the crystal in the
discussion. However starting from Sec.~IV radial ($ \xi $) and
azimuthal ($ \delta $) angles outside the nonlinear crystal are
naturally used in the presentation of experimental results.

\begin{figure}  % figs. 1a-h
 \resizebox{0.95\hsize}{!}{\includegraphics{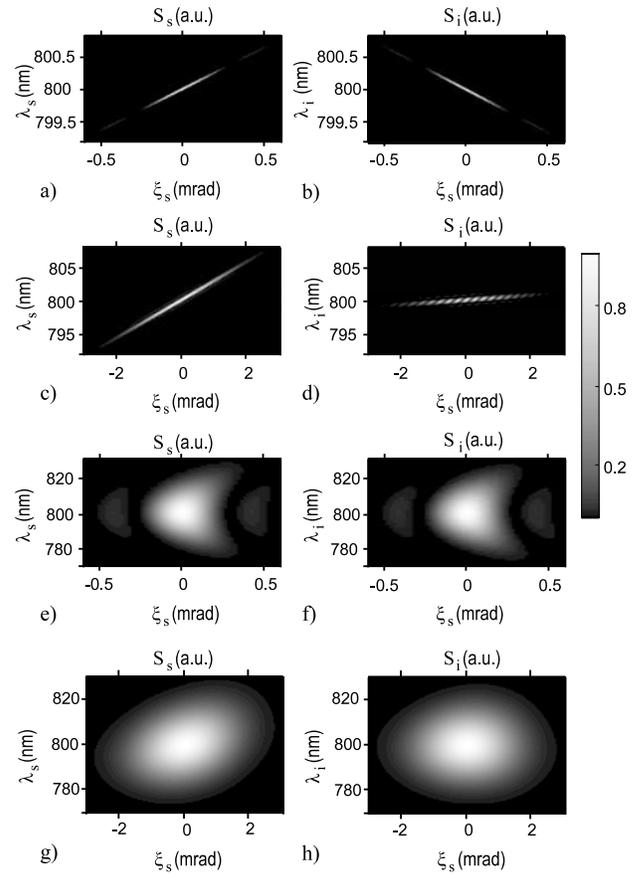}}
 \caption{Contour plots of signal- [$ S_s(\lambda_s) $] and idler-field
 [$ S_i(\lambda_i) $] intensity spectra
  as they depend
  on radial signal-field emission angle $ \xi_s $;
  idler-field emission angle $ \xi_i $ is fixed. Spectra are determined
  for cw plane-wave pumping (a, b), pulsed plane-wave pumping (c, d,
  $ \Delta\lambda_p = 2.8 $~nm), cw focused pumping (e, f, $ W_p^{0,f} = 20~\mu$m)
  and pulsed focused pumping (g, h, $ \Delta\lambda_p = 2.8 $~nm,
  $ W_p^{0,f} = W_{px}^{0,f} = W_{py}^{0,f} = 20~\mu$m) for $ L_z = 5 $~mm.}
 \label{fig1}
\end{figure}

In radial direction, crystal length and pump-field spectral
bandwidth as well as transverse pump-beam profile play a role. The
dependence of radial width of the correlation area on the crystal
length $ L_z $ emerges through the phase matching condition in the
$ z $ direction. This condition is mathematically described by the
expression $ {\rm sinc}(\Delta {\bf k}_z L_z/2) $ in
Eq.~(\ref{19}); $ \Delta {\bf k}_z = {\bf k}_{pz} - {\bf k}_{sz} -
{\bf k}_{iz} $. Actual radial width is determined by this
condition and conservation of energy ($ \omega_p = \omega_s +
\omega_i $). According to the formula in Eq.~(\ref{19}), the
longer the crystal, the smaller the radial width. Analytical
theory also predicts narrowing of the signal- and idler-field
spectra with an increasing crystal length. If pulsed pumping is
considered, the wider the pump-field spectrum, the greater the
radial width and also the greater the signal-field spectral width
(compare Figs.~\ref{fig1}c, d with Figs~\ref{fig1}a, b). This can
be understood as follows: more pump-field frequencies are present
in a wider pump-field spectrum and so more signal- and idler-field
frequencies are allowed to obey the phase-matching conditions in
the $ z $ direction and conservation of energy. In more detail and
following the graphs in Figs.~\ref{fig1}c and d, signal-field
photons with different wavelengths are emitted into different
radial emission angles $ \xi_s $. Superposition of photon fields
emitted into different radial emission angles $ \xi_s $ then
broadens the overall signal-field spectrum. It is important to
note that all idler-field photons have nearly the same wavelengths
which means that signal-field photons emitted into different
radial emission angles $ \xi_s $ use different wavelengths of the
pulsed-pump spectrum. The transverse pump-beam profile affects the
radial width through the phase-matching condition in the radial
plane. This radial phase matching condition is an additional
requirement that must be fulfilled by a generated photon pair.
Qualitatively, the more the pump beam is focused, the wider its
spatial spectrum in radial direction and so the weaker the radial
phase-matching condition. However, this dependence is quite small
in radial angles, as follows from the comparison of graphs in
Figs.~\ref{fig1}a, b and Figs.~\ref{fig1}e, f. On the other hand,
focusing of the pump beam leads to considerable broadening of the
signal- and idler-field spectra in all radial emission directions.
Finally, if a focused pulsed pump beam is assumed (see
Figs.~\ref{fig1}g, h), broadening of the correlation area in
radial direction as well as broadening of the overall signal- and
idler-field spectra is observed due to a final pump-field spectral
width. On the top, broadening of the signal- and idler-field
spectra corresponding to any radial emission angle $ \xi_s $
occurs as a consequence of pump-beam focusing. This behavior is
related to the fact that indexes of refraction of the interacting
fields are nearly constant inside the correlation area. We can say
in general, that spectral widths of the signal and idler fields
behave qualitatively in the same way as the radial width of
correlation area.

\begin{figure}  % figs. 2a-b
 \resizebox{0.95\hsize}{!}{\includegraphics{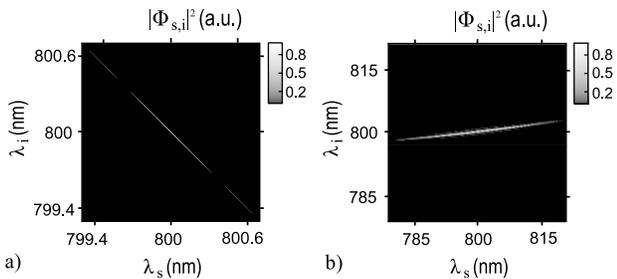}}
 \caption{Contour plots of squared modulus $ |\Phi_{s,i}|^2 $ of two-photon spectral amplitude
  for a) cw  and b) pulsed ($ \Delta\lambda_p = 8.5 $~nm) plane-wave pumping;
  $ L_z = 5 $~mm.}
 \label{fig2}
\end{figure}

Comparison of the signal- and idler-field spectra in
Figs.~\ref{fig1}c, d valid for pulsed pumping with those in
Figs.~\ref{fig1}a, b for cw pumping leads to a remarkable
observation. Photon pairs generated into different signal-photon
radial emission angles $ \xi_s $ use different pump-field
frequencies. There occurs spectral asymmetry between the signal
and idler fields that originates in different detection angles
considered; whereas the idler-field detection angle is fixed, the
angle of a signal-photon detection varies. This asymmetry
determines the preferred direction of the signal- and idler-field
frequency correlations as they are visible in the shape of squared
modulus $ |\Phi_{s,i}|^2 $ of two-photon spectral amplitude
introduced in Eq.~(\ref{10}) [a large signal-field detector is
assumed]. Contour plot of the squared modulus $ |\Phi_{s,i}|^2 $
of two-photon amplitude has a typical cigar shape. In cw case, the
main axis of this cigar is rotated by 45 degrees counter-clockwise
with respect to the $ \lambda_i $ axis (see Fig.~\ref{fig2}a) in
order to describe perfect frequency anti-correlation. If pulsed
pumping is taken into account, the cigar axis tends to rotates
clockwise; the broader the pump-field spectrum, the greater the
rotation angle. Even states with positively correlated signal- and
idler-field frequencies can be observed for sufficiently broad
pump-field spectra (see Fig.~\ref{fig2}b). We note that different
dispersion properties at different propagation angles have been
fully exploited in the method of achromatic phase matching that
allows to generate photon pairs with an arbitrary orientation of
the two-photon spectral amplitude
\cite{Torres2005,Torres2005a,Molina-Terriza2005}.

The azimuthal width of correlation area is determined
predominantly by a pump-beam transverse profile for geometric
reasons. To be more specific, it is the pump-beam spatial spectrum
in azimuthal direction that affects the azimuthal extension of the
correlation area through the phase-matching conditions in the
azimuthal direction. As material dispersion characteristics of the
crystal are rotationally symmetric with respect to the $ z $ axis
(signal and idler fields propagate as ordinary waves), the
azimuthal width of spatial pump-beam spectrum does not practically
influence spectral properties of the signal and idler fields.

\begin{figure}  % figs. 3a, b
 a) \resizebox{0.7\hsize}{0.55\hsize}{\includegraphics{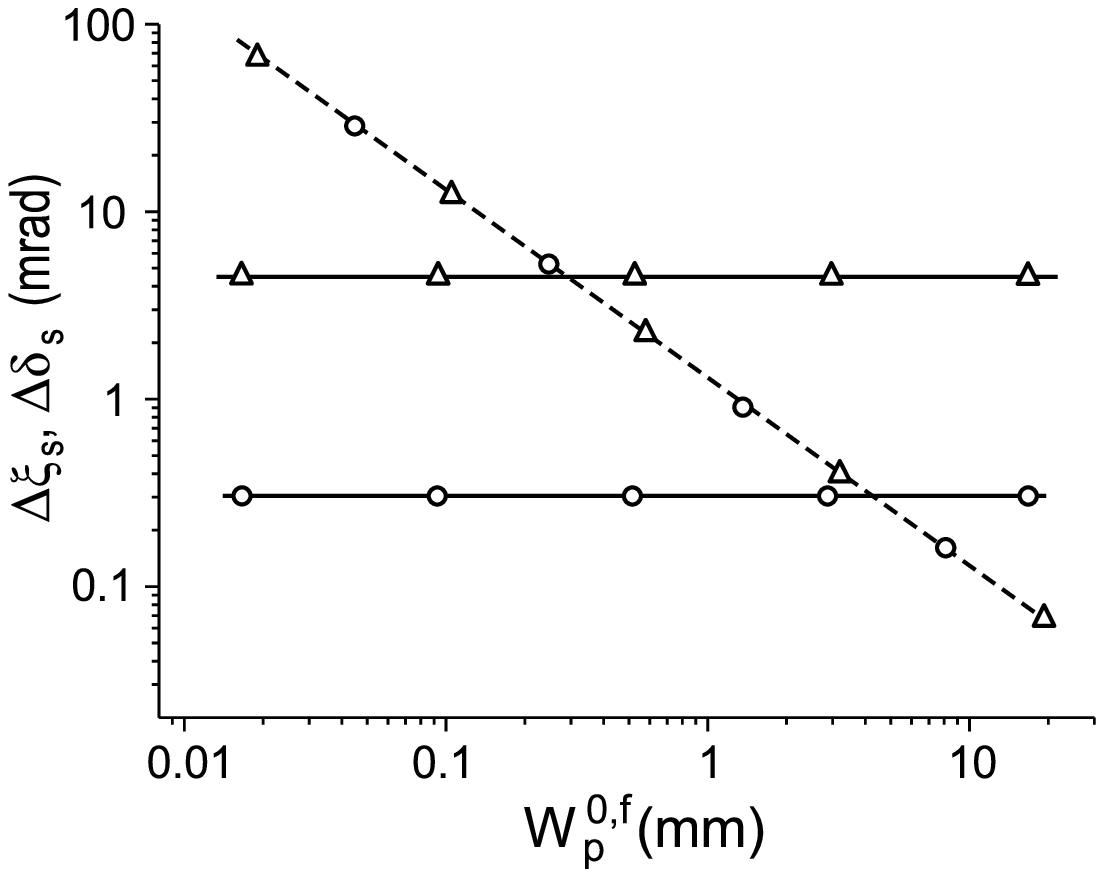}}

 b) \resizebox{0.7\hsize}{0.55\hsize}{\includegraphics{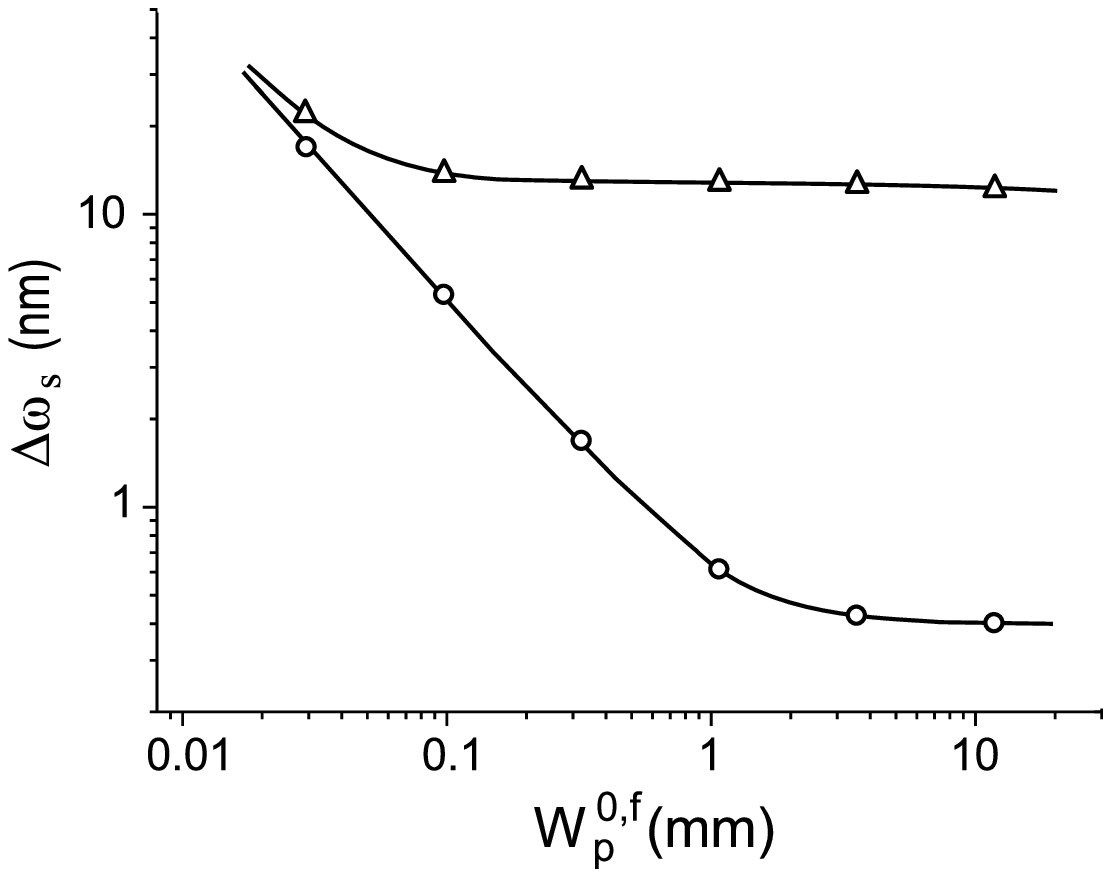}}

 \caption{a) Radial ($ \Delta\xi_s $, solid curves) and azimuthal ($
 \Delta\delta_s $, dashed curves) widths of correlation area and b) signal-field spectral
 width $ \Delta\omega_s $ as they depend on width $
 W_p^{0,f} $ of the pump-beam waist for pulsed ($ \Delta\lambda_p = 5 $~nm, triangles)
 and cw ($ \Delta\lambda_p = 0.03 $~nm, circles) pumping; $ L_z = 5
 $~mm. Logarithmic scales on the $ x $ and $ y $ axes are used.}
 \label{fig3}
\end{figure}

We illustrate the dependence of correlation area on pump-beam
focusing using a 5~mm long crystal and both cw and pulsed pumping
in Fig.~\ref{fig3}. We can see in Fig.~\ref{fig3}a that the
signal-field azimuthal width $ \Delta\delta_s $ is inversely
proportional to the width $ W_p^{0,f} $ (full-width at $ 1/e $ of
the maximum; $ W_p^{0,f} \equiv W_{px}^{0,f} = W_{py}^{0,f} $) of
the pump-beam waist whereas the radial width $ \Delta\xi_s $ does
not practically depend on the width $ W_p^{0,f} $ of the pump-beam
waist. This is caused by the fact that the phase-matching
condition in the $ z $ direction is much stronger than that in
radial direction for a 5-mm long crystal and so the radial width $
\Delta\xi_s $ is sensitive only to the pump-field spectral width
in this case. Pulsed pumping gives a broader correlation area in
radial direction as well as broader signal-field spectrum compared
to cw case (see Fig.~\ref{fig3}b). Increasing pump-beam focusing
releases phase-matching conditions and naturally leads to a
broader signal-field spectrum.

\begin{figure}   % figs. 4a,b
 a) \resizebox{0.7\hsize}{0.50\hsize}{\includegraphics{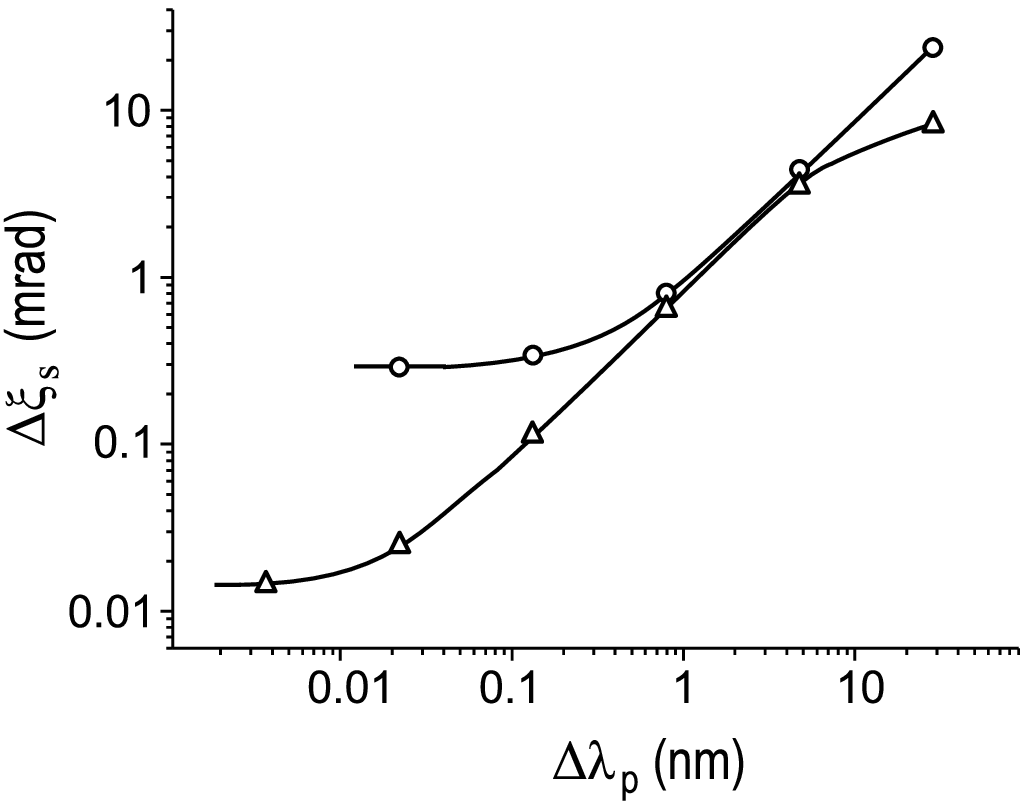}}

 b) \resizebox{0.7\hsize}{0.50\hsize}{\includegraphics{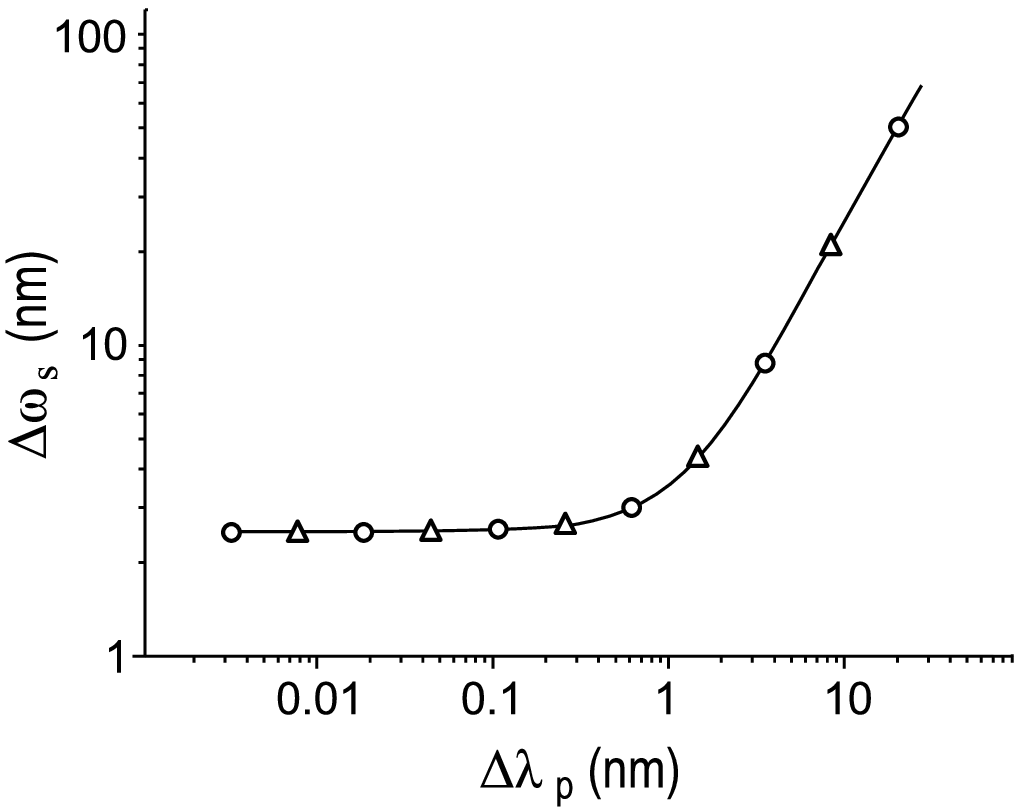}}

 \caption{a) Radial width $ \Delta\xi_s $ of correlation area
  and b) signal-field spectral width $ \Delta\omega_s $ as
  functions of pump-field spectral width $ \Delta\lambda_p $ for
  a 5-mm (circles) and 10-cm (triangles) long crystal assuming a
  focused pump beam; $ W_{p}^{0,f} = 200\;\;\mu $m. Logarithmic scales
  on the $ x $ and $ y $ axes are used.}
\label{fig4}
\end{figure}

Contrary to the azimuthal width, the radial width $ \Delta\xi_s $
depends on the pump-field spectral width $ \Delta\lambda_p $. The
larger the pump-field spectral width $ \Delta\lambda_p $ the
greater the radial width $ \Delta\xi_s $ and also the greater the
signal-field spectral width $ \Delta\omega_s $, as documented in
Fig.~\ref{fig4} for a focused pump beam. We can also see in
Fig.~\ref{fig4}a that the radial width $ \Delta\xi_s $ reaches a
constant value for sufficiently narrow pump-field spectra. This
value is determined by the phase-matching condition in the $ z $
direction for the central pump-field frequency $ \omega_p^0 $ and
so depends on the crystal length $ L_z $ (together with material
dispersion properties of the crystal). The longer the crystal the
smaller the radial width $ \Delta\xi_s $.

\begin{figure}   % figs. 5a,b
 a) \resizebox{0.7\hsize}{0.5\hsize}{\includegraphics{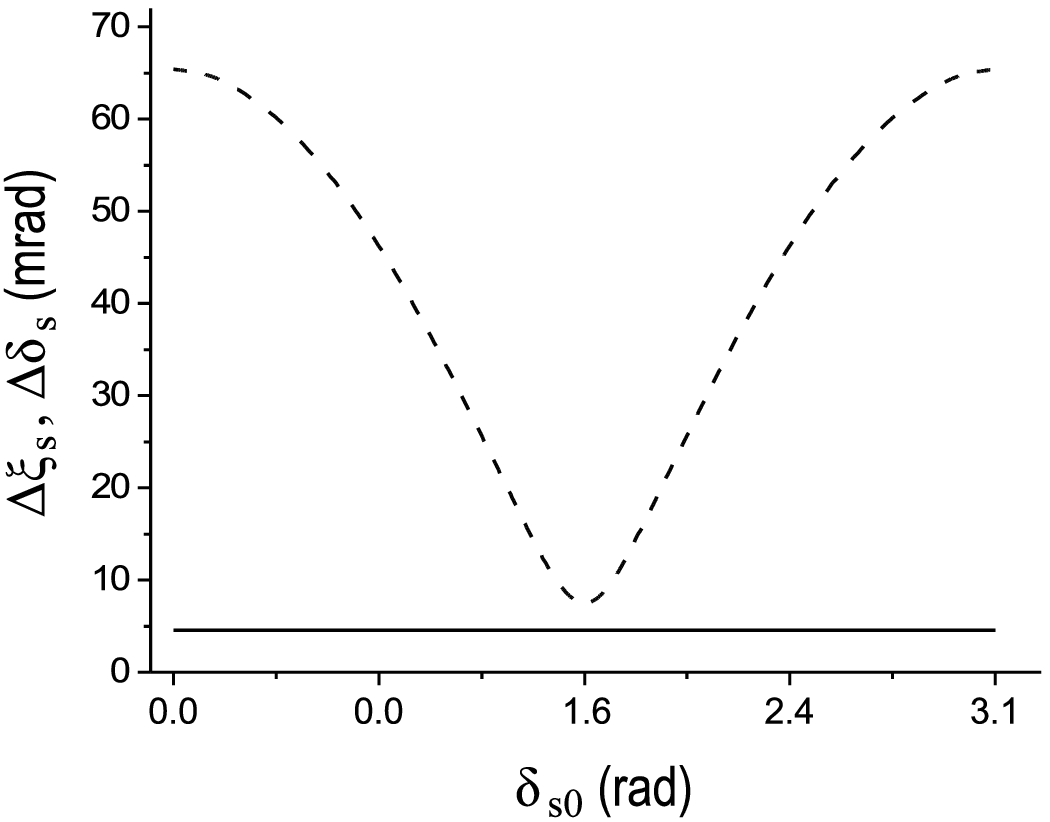}}

 b) \resizebox{0.7\hsize}{0.5\hsize}{\includegraphics{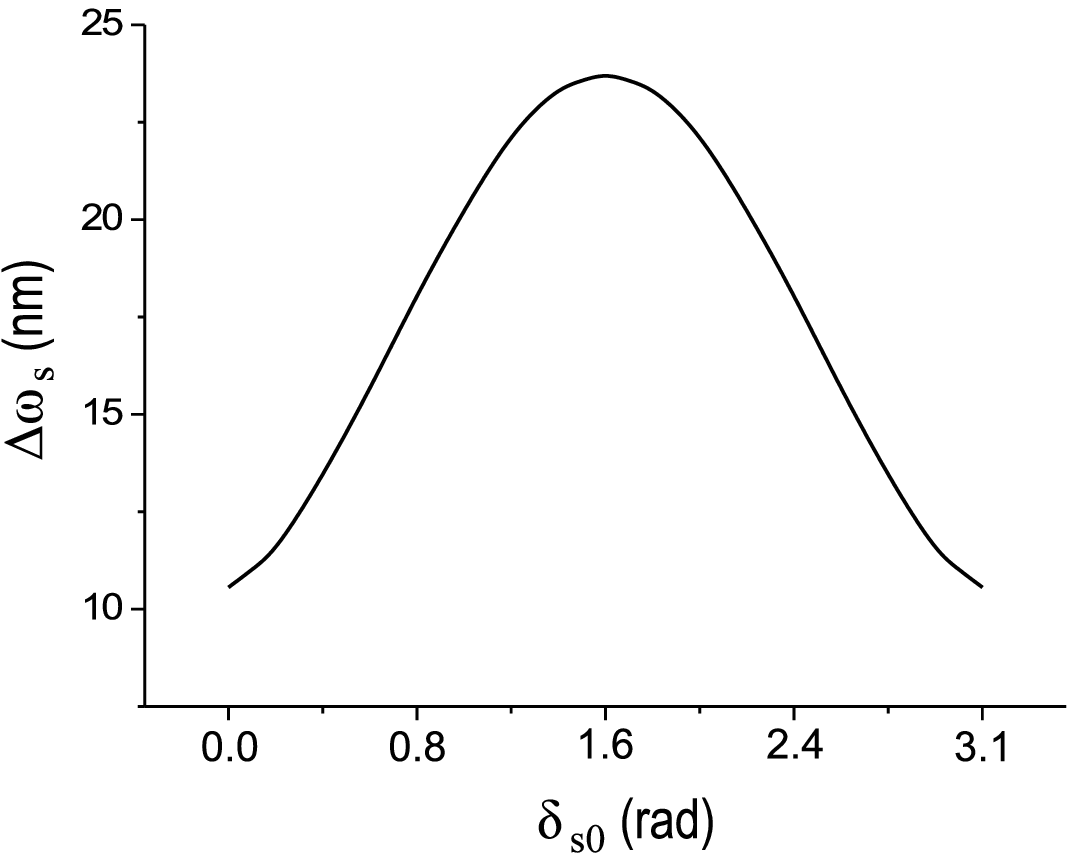}}

 \caption{a) Radial ($ \Delta\xi_s $, solid curve) and azimuthal ($
 \Delta\delta_s $, dashed curve) widths of correlation area and b) signal-field spectral
 width $ \Delta\omega_s $ as they depend on central azimuthal signal-field emission angle
 $ \delta_{s0} $ for a highly elliptic pulsed pump beam ($ W_{px}^{0,f} = 200\;\;\mu $m,
 $ W_{py}^{0,f} = 20\;\;\mu $m, $ \Delta\lambda_p = 5 $~nm); $ L_z = 5 $~mm.}
\label{fig5}
\end{figure}

The above described dependencies allow to generate photon pairs
with highly elliptic profiles of the correlation area provided
that the pump-beam profile in the transverse plane is highly
elliptic. As an example, we consider a pump beam having $
W_{py}^{0,f}/W_{px}^{0,f} = 10 $. The dependence of the radial ($
\Delta\xi_s $) and azimuthal ($ \Delta\delta_s $) widths and
signal-field spectral width $ \Delta\omega_s $ on the central
azimuthal signal-photon emission angle $ \delta_{s0} $ is shown in
Fig.~\ref{fig5} in this case. Whereas the radial and azimuthal
widths are comparable for the azimuthal signal-field emission
angle $ \delta_{s0} = \pi/2 $, their ratio $
\Delta\delta_s/\Delta\xi_s $ equals approx. 20 for $ \delta_{s0} =
0 $. Focusing the pump beam from 200~$ \mu $m to 20~$ \mu $m in
radial direction results in doubling the signal-field spectral
width $ \Delta\omega_s $ as documented in Fig.~\ref{fig5}b (see
also Figs.~\ref{fig1}a and e).

\section{Experimental setup}

\begin{figure}  % fig. 6
 \scalebox{0.75}{\includegraphics{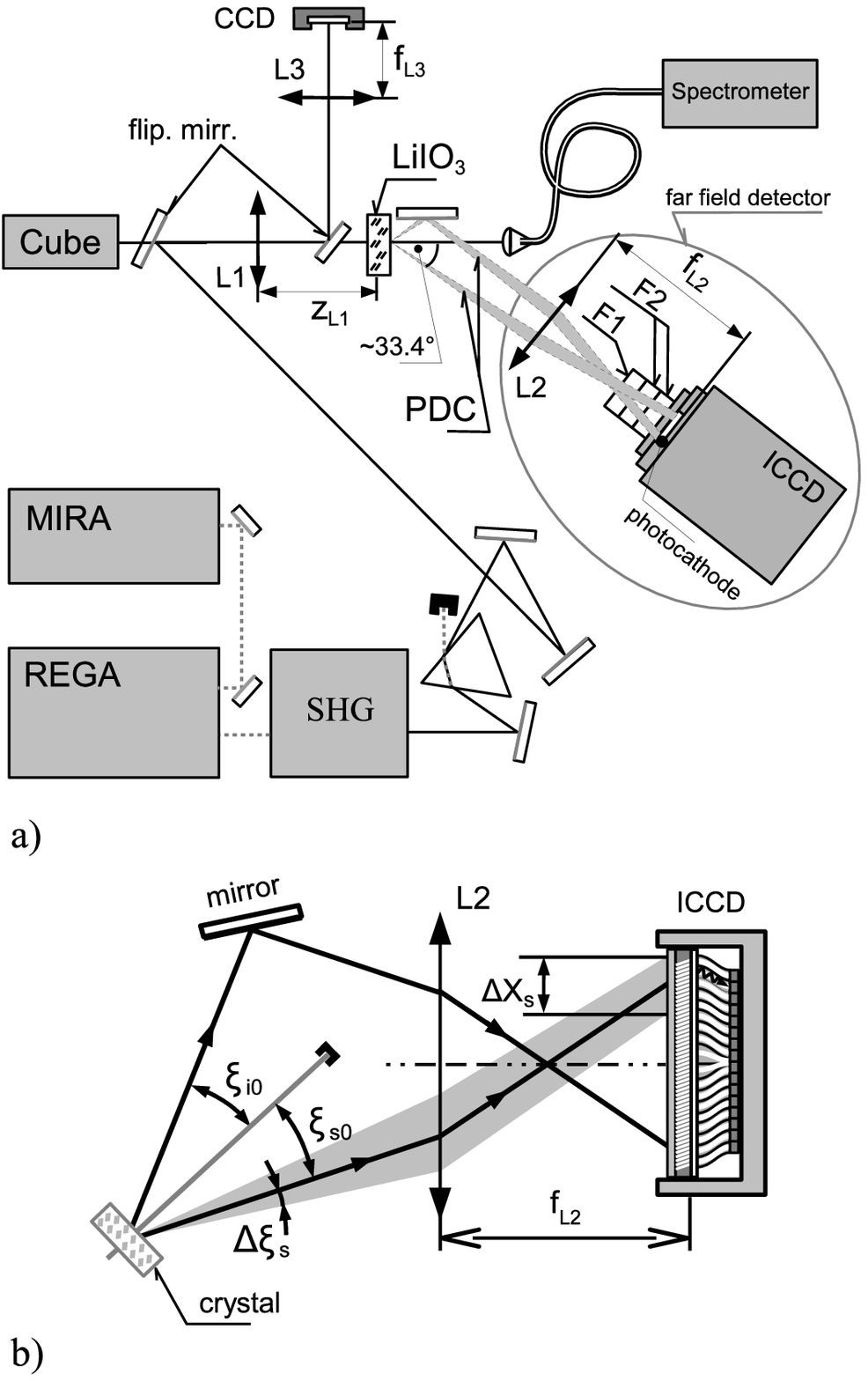}}
 \caption{Experimental setup used for the determination of
 angular widths: a) Entire setup that includes both
 cw and pulsed pumping as well as pump-beam diagnostics (for
 more details, see the text). b) Detail of the setup showing paths
 of the signal and idler beams.}
\label{fig6}
\end{figure}

We have used a negative uniaxial crystal made of LiIO$_3$ cut for
non-critical phase matching, i.e. the optical axis was
perpendicular to the pump-beam propagation direction. We have
considered crystals of two different lengths ($L_z$=2~mm and 5~mm)
pumped both by cw and pulsed lasers. As for cw pumping, a
semiconductor laser Cube 405 (Coherent) delivered 31.6~mW at
405~nm and with spectral bandwidth $\Delta \lambda_p= 1.7 $~nm.
The second-harmonic field of an amplified femtosecond Ti:sapphire
system (Mira+RegA, Coherent) providing pulses at 800~nm and
$\sim$250~fs long was used in the pulsed regime. The mean SHG
power was 2.5~mW at the crystal input for a repetition rate of
11~kHz. Spectral bandwidth was adjusted between 4.8 and 7.4~nm by
fine tuning of the SHG process. A dispersion prism was used to
separate the fundamental and SHG beams (for details, see
Fig.~\ref{fig6}).

Transverse profile of the pump beam and its divergence were
controlled by changing the focus length of converging lens L1 or
using a beam expander (BE2X, Thorlabs). The used focal lengths
$f_{L1}$ of lens $ L1 $ laid in the interval from 30 to 75~cm. As
we wanted the pump beam to be as homogeneous as possible along the
$ z $ axis, we chose the distance $z_{L1}$ between the lens L1 and
the nonlinear crystal such that the beam waist was placed far
behind the crystal, i.e. $z_{L1}<f_{L1}$. Spatial spectrum of the
pump beam in the transverse plane as a very important parameter in
our experiment was measured by a CCD camera (Lu085M, Lumenera)
placed at the focal plane of a converging lens L3. Spatial spectra
in horizontal and vertical directions were determined as marginal
spectra and parameters $ \tilde{W}_{px} $ and $ \tilde{W}_{py} $
characterizing their widths were found after fitting the
experimental data. A fiber-optic spectrometer (HR4000CG-UV-NIR,
Ocean Optics) was used to obtain the pump-beam temporal spectrum
after propagation through the nonlinear crystal.

The experiment was done with photon pairs degenerate in
frequencies ($\lambda _{s0}=\lambda _{i0} = 800$~nm) and emitted
in opposite parts of a cone layer (the central radial emission
angle was 33.4~deg behind the crystal). As shown in
Fig.~\ref{fig6}b the signal beam was captured directly by a
detector whereas the idler beam propagated to the detector after
being reflected on a high-reflectivity mirror. Both beams were
detected on a photocathode of an iCCD camera with image
intensifier (PI-MAX:512-HQ, Princeton Instruments). Before
detection, both beams were transformed using a converging lens L2,
one narrow-bandwidth and two high-pass edge filters. The geometry
of the setup was chosen such that the lens L2 mapped the signal
and idler photon emission angles to positions at the photocathode;
the photocathode was placed in the focal plane of lens $ L2 $. For
convenience, lenses L2 with different focal lengths ($f_{L2}$=
12.5, 15, and 25~cm) were used. The applied bandwidth filter was
11~nm wide and centered at 800~nm. Edge filters (Andover,
ANDV7862) had high transmittances at 800~nm (98\%) and blocked
wavelengths below 666~nm.

\begin{figure}   % fig. 7
 \scalebox{0.6}{\includegraphics{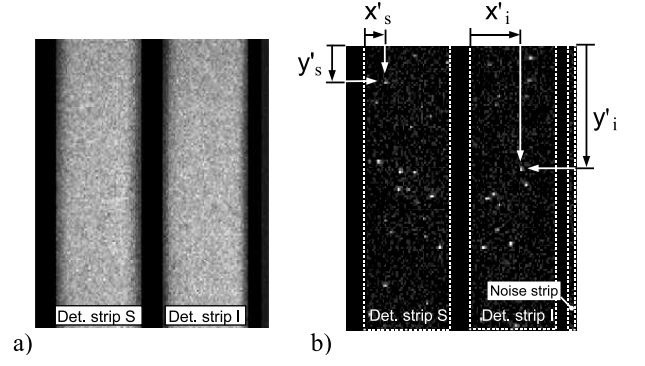}}
 \caption{Photocathode with registered photons after a) illumination
 by light coming from 20 000 consecutive pump pulses, b) one pump pulse.
 The signal and idler strips image small sections of the cone layer and are
 slightly curved. The curvatures are oriented in the same
 sense in both strips because the idler beam is reflected on a
 mirror. } \label{fig7}
\end{figure}

An active area of the photocathode in the form of a rectangular
12.36~mm wide (see Fig.~\ref{fig7}) was divided into $512\times
512$ pixels. Spatial resolution of the camera was 38~$\mu$m (FWHM)
and its main limitation came from imperfect contrast transfer in
the image intensifier. In order to make data acquisition faster
the resolution was further decreased by grouping $4\times 4$ or
$8\times 8$ pixels into one super-pixel in the hardware of the
camera. Consequently, several tens of camera frames were captured
in one second. The overall quantum detection efficiency including
components between the crystal and photocathode was 7\%, as
derived from covariance of the signal and idler photon numbers.
Widths of the signal and idler strips are given by the bandwidth
filter and lens L2 focal length. As for timing, a 10~ns long gate
of the camera was used synchronously with laser pulses. In cw
case, a 2~$\mu$s long gate was applied together with internal
triggering. This timing together with appropriate pump-field
intensities assured that the probability of detecting two photons
in a single super-pixel was negligible. In other words, the number
of detection events divided by quantum detection efficiency had to
be much lower than the number of super-pixels.

Also the level of noise was monitored in the third narrow strip;
1.82\% of detection events came from noise. Detailed analysis has
shown that 90~\% of noise photons were red photons originating
from fluorescence inside the crystal. Scattered pump photons
contributed by 8.4~\% and only 1.6\% of noise counts were dark
counts of the iCCD camera.

The experimental signal-idler correlation functions $ g_x $ and $
g_y $ in the transverse plane described by horizontal ($x'$) and
vertical ($y'$) coordinates of the reference system in this plane
have been determined after processing many experimental frames.
The formula for the determination of correlation function $ g_x $
can be written as follows (see also Fig.~\ref{fig7}b):
\begin{equation}  % 20
 g_x( x'_s ,x'_i ) =
  \sum_{p = 1}^N {} \sum_{m = 1}^{M_p } {} \sum_{l = 1}^{L_p }
  \delta \left( {x'}_{s}^{pm}  - x'_s
  \right)\delta \left( {x'}_{i}^{pl}  - x'_i \right).
\label{20}
\end{equation}
In Eq.~(\ref{20}), $p$ indexes frames ($N$ gives the number of
frames) and $m$ ($l$) counts signal (idler) detection events [up
to $M_p$ ($L_p$) in the $p$-th frame]. Symbol $ {x'}_s^{pl}$ ($
{x'}_i^{pl}$) denotes horizontal position of the $l$-th detection
in the signal (idler) strip of the $p$-th frame. Correlations in
the vertical direction given by the correlation function $ g_y $
can be determined similarly. The formula in Eq.~(\ref{20}) takes
into account all possible combinations of pairwise detection
events. Only some of them correspond to detection of both photons
from one pair. The remaining combinations are artificial in the
sense that they do not correspond to detection of a photon pair.
This poses the following restriction to the method. The number of
artificial combinations that occur at random positions has to be
large enough in order to create a plateau in a 2D graph of
correlation function $ g_x( x'_s,x'_i ) $. Real detections of
photon pairs are then visible on the top of this plateau (see
Fig.~\ref{fig9} later).

\begin{figure}   % fig. 8
 \scalebox{0.7}{\includegraphics{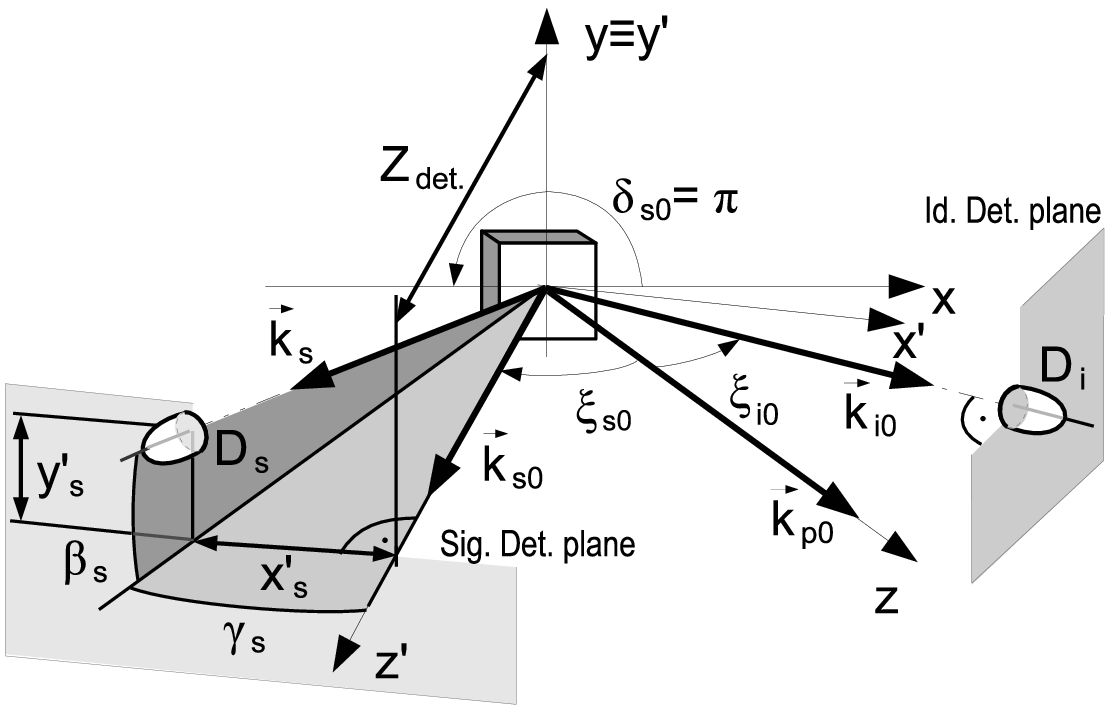}}
 \caption{Sketch showing the geometry of signal and idler beams.
  The photon emission direction is described by radial ($ \xi $) and azimuthal
  ($ \delta $) emission angles. In detector plane, cartesian
  coordinates $ x' $ and $ y' $ are useful. Photons propagation
  directions are then conveniently parameterized by angles $ \beta
  $ and $ \gamma $.}
\label{fig8}
\end{figure}

Cartesian coordinates $ x'_j $ and $ y'_j $, $ j=s,i $, in the
transverse plane can be conveniently transformed into angles $
\beta_j $ and $ \gamma_j $ measured from the middle ($ x'^{\rm
cent}_j $, $ y'^{\rm cent}_j $) of the $ j $th strip and defined
in Fig.~\ref{fig8} using the formulas:
\begin{eqnarray}      % 21
 \gamma_j &=& \arctan\left[(x'_j- x'^{\rm
  cent}_j) /f_{L2}\right] , \nonumber \\
 \beta_j &=& \arctan\left[ (y'_j-y'^{\rm
  cent}_j) /f_{L2} \cos(\gamma_j)
 \right], \hspace{0.3cm} j=s,i; \nonumber \\
 & &
\label{21}
\end{eqnarray}
$ f_{L2} $ means the focal length of lens $ L2 $. Angles $ \beta_j
$ and $ \gamma_j $ are related to radial and azimuthal angles $
\xi_j $ and $ \delta_j $ by the following transformation:
\begin{eqnarray}   % 22
 \beta_j &=& \arcsin\left[ \sin(\xi_j)\sin(\delta_j)\right] ,
  \nonumber \\
 \gamma_j &=& \arctan \left[ \tan(\xi_j) \cos(\delta_j) \right] -
  \xi_{j,{\rm det}} , \hspace{0.2cm} j=s,i,
\label{22}
\end{eqnarray}
where the radial angle $ \xi_{j,{\rm det}} $ describes the
position of a detector in beam $ j $.

\section{Experimental determination of parameters of correlation area}

\begin{figure}   % fig. 9
 \scalebox{0.7}{\includegraphics{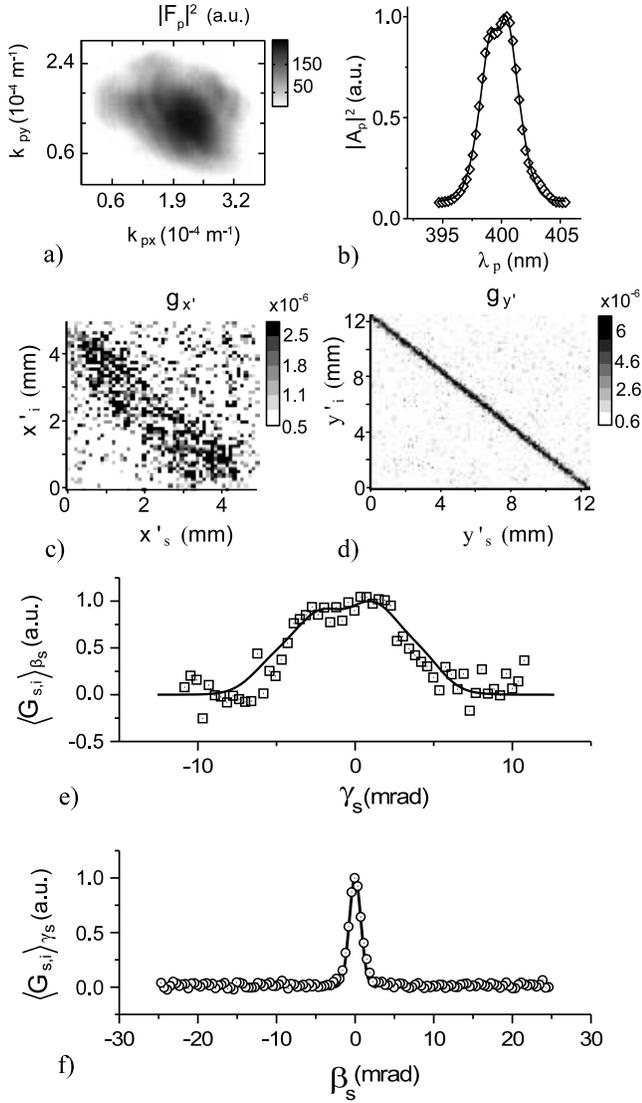}}
\caption{Typical measurement of a correlation area for pulsed
 pumping having 327,600 frames; $L_z=5$~mm. a) Spatial
 spectrum of the pump beam determined in the focal plane of lens L3.
 b) Temporal intensity spectrum of the pump beam as determined by a spectrometer
  (diamonds), solid line represents a multi-peak Gaussian fit.
 c), d) Experimental histograms $ g_x(x'_s,x'_i) $ (c) and
 $ g_y(y'_s,y'_i) $ (d). e), f) Experimental radial ($ \langle G_{s,i}\rangle_{\beta_s} $,
 e) and azimuthal ($ \langle G_{s,i}\rangle_{\gamma_s} $, f)
 cross-sections of
 the correlation area (rectangles, circles) together with theoretical
 predictions (solid lines).}
\label{fig9}
\end{figure}

In the experiment, spatial and temporal spectra of the pump beam
have been characterized first. Typical results are shown in
Figs.~\ref{fig9}a and b and have been used in the model for the
determination of expected parameters of the correlation area. The
correlation area, or more specifically its radial and angular
profiles, have been characterized using histograms $
g_x(x'_s,x'_i) $ and $ g_y(y'_s,y'_i) $. Histogram $
g_x(x'_s,x'_i) $ [$ g_y(y'_s,y'_i) $] gives the number of paired
detections with a signal photon detected at position $ x'_s $ [$
y'_s $] together with an idler photon registered at position $
x'_i $ [$ y'_i $]. These histograms usually contain experimental
data from several hundreds of thousands of frames. As graphs in
Figs.~\ref{fig9}c and d show detections of correlated photon pairs
lead to higher values in histograms $ g_x $ and $ g_y $ around
diagonals going from upper-left to lower-right corners of the
plots. Finite spreads of these diagonals have their origin in
non-perfect phase matching and can be characterized by their
widths $ \Delta x'_s $ and $ \Delta y'_s $. Or more conveniently
by uncertainties in the determination of angles $ \beta_s $ and $
\gamma_s $; $\Delta \beta_{s} \approx \Delta y'_{s}/f_{L2}$ and
$\Delta \gamma_{s} \approx \Delta x'_{s}/f_{L2}$. As detailed
inspection of the histogram $ g_x $ ($ g_y $) in Fig.~\ref{fig9}c
(d) has shown, cuts of this histogram along the lines with
constant values of $ x'_i $ ($ y'_i $) do not depend on the value
of $ x'_i $ ($ y'_i $). This reflects the fact that idler photons
detected at different positions inside the investigated area on
the photocathode have identical (signal-photon) correlation areas.
This allows us to combine the data obtained for idler photons
detected at different positions together and increase the
measurement precision this way. This approach thus provides the
radial cross-section $ \langle G_{s,i}\rangle_{\beta_s} $ of the
correlation area along the radial angle $ \gamma_s $ as a mean
value over all possible values of the signal-field azimuthal angle
$ \beta_s $. Moreover, consideration of different idler-photon
detection positions means averaging over the angles $ \gamma_i $
and $ \beta_i $. The averaging is indicated by symbol $ \langle
\rangle $. Mathematically, the radial cross-section $ \langle
G_{s,i}\rangle_{\beta_s} $ expressed in the coordinate $ x'_s $
can be derived along the formula
\begin{equation}   % 23
 \langle G_{s,i}\rangle_{\beta_s}(x'_s) = \sum_{x'_i} g \left[
  x'_s - x'^{\rm mid}_s(x'_i),x'_i\right] ,
\label{23}
\end{equation}
where the function $ x'^{\rm mid}_s(x'_i) $ gives the central
position (given as a locus) of the cut of the histogram $
g(x'_s,x'_i) $ for a fixed value of the coordinate $ x'_i $. In
the theory, the radial cross-section $ \langle
G_{s,i}\rangle_{\beta_s} $ is determined using the fourth-order
correlation function $ G_{s,i} $ written in Eq.~(\ref{7}),
substitution of angles $ \xi_s $, $ \delta_s $, $ \xi_i $, and $
\delta_i $ by angles $ \gamma_s $, $ \beta_s $, $ \gamma_i $, and
$ \beta_i $ [inverse transformation to that in Eq.~(\ref{22})] and
finally integration over the angles $ \beta_s $, $ \gamma_i $, and
$ \beta_i $. Similarly, the azimuthal cross-section $ \langle
G_{s,i}\rangle_{\gamma_s} $ of the correlation area along the
azimuthal angle $ \beta_s $ arises after averaging over the angles
$ \gamma_s $, $ \gamma_i $, and $ \beta_i $ and can be determined
by a formula analogous to that given in Eq.~(\ref{23}). The radial
and azimuthal cross-sections $ \langle G_{s,i}\rangle_{\beta_s} $
and $ \langle G_{s,i}\rangle_{\gamma_s} $ corresponding to the
pump beam with characteristics defined in Figs.~\ref{fig9}a and b
are plotted in Figs.~\ref{fig9}e and f. Solid lines in
Figs.~\ref{fig9}e and f refer to the results of numerical model
and are in a good agreement with the experimental data.

\begin{figure}   % fig. 10
 \scalebox{0.65}{\includegraphics{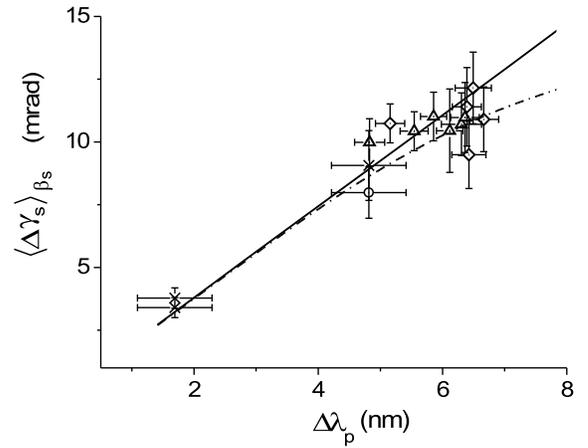}}
 \caption{Radial width
 $\langle\Delta \gamma_{s}\rangle_{\beta_s}$ as a function of
 pump-beam spectral width $ \Delta\lambda_p $. Experimental points
 have been obtained for $ L_{z}=5 $~mm, $ W_{px}^{0,f} > 140~\mu $m (triangles),
 $ L_{z}=5 $~mm, $ W_{px}^{0,f} < 140~\mu $m (diamonds),
 $ L_{z}=2 $~mm, $ W_{px}^{0,f} > 140~\mu $m (crosses), and
 $ L_{z}=2 $~mm, $ W_{px}^{0,f} < 140~\mu $m (circles) both for cw and pulsed
 pumping. The theoretical model gives the same dependence for both crystal lengths $ L_z $
 in cases without (solid curve) as well as with (dashed curve, 11-nm wide) spectral filters.}
\label{fig10}
\end{figure}

The radial width $\langle\Delta \gamma_{s}\rangle_{\beta_s}$
(measured as full-width at $ 1/e $ of the maximum) of radial
cross-section $ \langle G_{s,i}\rangle_{\beta_s} $ depends mainly
on the pump-beam spectral width $\Delta \lambda_{p}$. It holds
that the greater the pump-beam width $\Delta \lambda_{p}$ the
larger the radial width $\langle\Delta
\gamma_{s}\rangle_{\beta_s}$ as documented in Fig.~\ref{fig10} for
crystals 2- and 5-mm long. In the experiment, 11-nm wide frequency
filters have been applied to cut noise. However, certain amount of
photons comprising a photon pair has also been blocked. According
to the theoretical model, this has also resulted in a small
narrowing of the radial cut of the correlation area (compare solid
and dashed curves in Fig.~\ref{fig10}). The theoretical curve in
Fig.~\ref{fig10} has been experimentally confirmed for several
values of the width $W_{px}^{0,f}$ of pump-beam waist both for cw
and pulsed pumping.

\begin{figure}   % fig. 11
 \scalebox{0.6}{\includegraphics{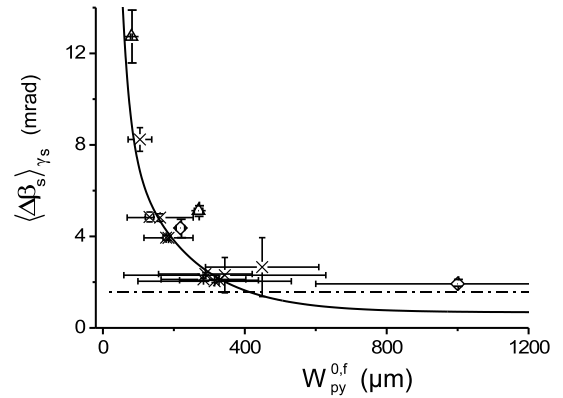}}
 \caption{Angular width
 $\langle\Delta \beta_{s}\rangle_{\gamma_s}$ as it depends on width
 $W_{py}^{0,f}$ of the pump-beam waist for pulsed (crosses and triangles) and cw
 (diamonds) pumping. 2-mm (triangles and diamonds) and 5-mm
 (crosses) long crystals were used in experiment. The theoretical model predicts the
 same dependence for both crystals (solid curve).
 The dashed horizontal line indicates the
 resolution limit given by the camera super-pixel size.}
\label{fig11}
\end{figure}

On the other hand and in our geometry, it is the width $
W_{py}^{0,f} $ of the pump-beam waist that determines the angular
width $\langle\Delta \beta_{s}\rangle_{\gamma_s}$ of angular
cross-section $ \langle G_{s,i}\rangle_{\gamma_s} $. Predictions
of the model for 2- and 5-mm long crystals are shown in
Fig.~\ref{fig11} by a solid curve. This curve has been checked
experimentally for several values of the width $ W_{py}^{0,f} $ of
the pump-beam waist both for cw and pulsed pumping. We note that
these curves do not depend on the pump-beam spectral width $
\Delta\lambda_p $. We can see in Fig.~\ref{fig11} that the
measured points agree with the theoretical curve for smaller
values of the width $ W_{py}^{0,f} $. Larger values of the width $
W_{py}^{0,f} $ lead to small angular widths $\langle\Delta
\beta_{s}\rangle_{\gamma_s}$ that could not be correctly measured
because of the limited spatial resolution of the iCCD camera.

\section{Engineering the shape of a correlation area}

\begin{figure}   % fig. 12
 \scalebox{0.65}{\includegraphics{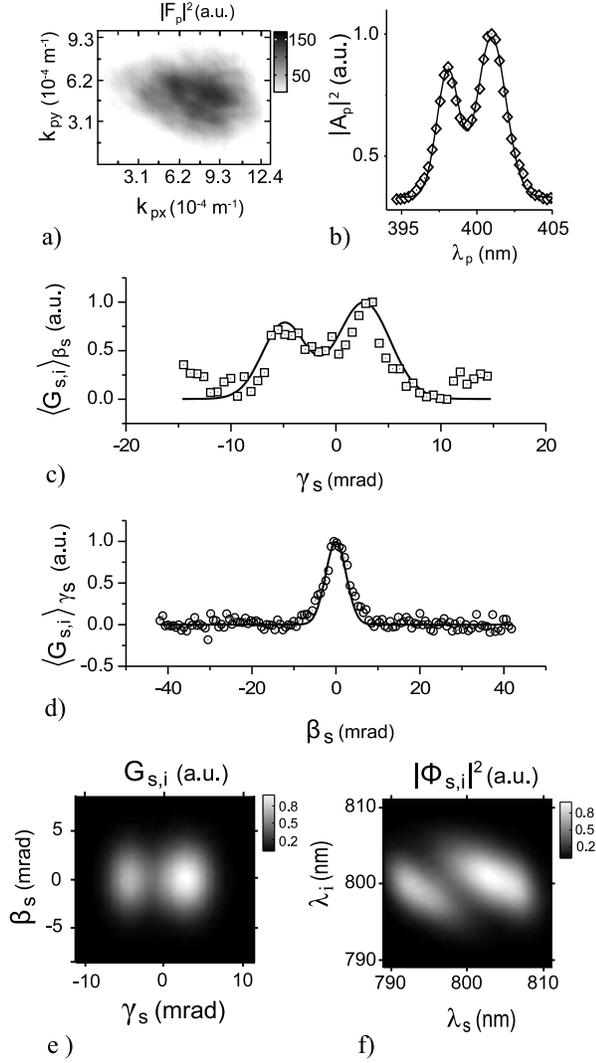}}
 \caption{Determination of a correlation area for pulsed
 pumping composed of two spectral peaks; $L_z=5$~mm. a) Spatial
 spectrum of the pump beam. b) Temporal pump-field intensity spectrum
 (experimental points are indicated by diamonds, solid line
 represents a multi-peak Gaussian fit).
 c) ,d) Experimental radial ($ \langle G_{s,i}\rangle_{\beta_s} $, c)
 and azimuthal ($ \langle G_{s,i}\rangle_{\gamma_s} $, d) cross-sections of
 the correlation area (rectangles and circles) together with theoretical
 predictions (solid line). e) Contour plot of the theoretical correlation function
 $ G_{s,i} $. f) Contour plot of the squared modulus $ |\Phi_{s,i}|^2 $ of the theoretical
 two-photon spectral amplitude.}
\label{fig12}
\end{figure}

As the above results have shown parameters of a correlation area
can be efficiently controlled using pump-beam parameters, namely
temporal spectrum and transverse profile. Even the shape of
correlation area can be considerably modified. Splitting of the
correlation area into two parts that occurs as a consequence of
splitting of the pump-field temporal spectrum can serve as an
example. Using our femtosecond pump system, we were able to
experimentally confirm this behavior. We have generated a pump
beam with the spatial spectrum given in Fig.~\ref{fig12}a. Its
temporal spectrum containing two peaks as was acquired by a
spectrometer is plotted in Fig.~\ref{fig12}b. The experimental
radial width $\langle\Delta \gamma_{s}\rangle_{\beta_s}$ of
cross-section $ \langle G_{s,i}\rangle_{\beta_s} $ given in
Fig.~\ref{fig12}c shows that the two-peak structure of the
pump-field spectrum resulted in splitting of the correlation area
into two parts. On the other hand and in agreement with the
theory, the angular cross-section $ \langle
G_{s,i}\rangle_{\gamma_s} $ was not affected by the pump-field
spectral splitting (see Fig.~\ref{fig12}d). For comparison, the
theoretical profile of the correlation area given by the
correlation function $ G_{s,i} $ and appropriate for the pump-beam
parameters given in Figs.~~\ref{fig12}a and b is plotted in
Fig.~\ref{fig12}e. It indicates a good agreement of the model with
experimental data. Moreover, the squared modulus $ |\Phi_{s,i}|^2
$ of theoretical two-photon spectral amplitude reveals that
splitting of the correlation area is accompanied by splitting of
the signal-field spectrum (see Fig.~\ref{fig12}f).

\section{Conclusions}

We have developed a method for the determination of profiles of a
correlation area using an intensified CCD camera. Single detection
events in many experimental frames are processed and provide
histograms from which cross-sections of the correlation area can
be recovered. This method has been used for investigations of the
dependence of parameters of the correlation area on pump-beam
characteristics and crystal length. The experimentally obtained
curves have been successfully compared with a theoretical model
giving fourth-order correlation functions. Radial profile of the
correlation area depends mainly on pump-field spectrum and crystal
length. On the other hand, azimuthal profile of the correlation
area is sensitive only to the transverse profile of the pump beam.
Splitting of the correlation area caused by a two-peak structure
of the pump-field spectrum has also been experimentally observed.

\acknowledgments This research has been supported by the projects
IAA100100713 of GA AV \v{C}R, 1M06002 and COST OC 09026 of the
Ministry of Education of the Czech Republic.


\begin{thebibliography}{99}

\bibitem{Giallorenziho1968} T. G. Giallorenziho and C. L. Tang, Phys. Rev. 166, 225 (1968).
\bibitem{Hong1985} C. K. Hong and L. Mandel, Phys. Rev. A 31, 2409 (1985).
\bibitem{Wang1991} L. J. Wang, X. Y. Zou, and L. Mandel, Phys. Rev. A 44, 4614 (1991).
\bibitem{Grayson1994} T. P. Grayson and G. A. Barbosa, Phys. Rev. A 49, 2948 (1994)
\bibitem{Steuernagel1998} O. Steuernagel and H. Rabitz, Opt. Com. 154, 285 (1998).
\bibitem{Keller1997} T. E. Keller and M. H. Rubin, Phys. Rev. A. 56, 1534 (1997).
\bibitem{DiGiuseppe1997} G. Di Giuseppe, L. Haiberger, F. De
 Martini, and A. V. Sergienko, Phys. Rev. A 56, R21 (1997).
\bibitem{Grice1997} W. P. Grice and I. A. Walmsley,
 Phys. Rev. A 56, 1627 (1997).
\bibitem{PerinaJr1999} J. Pe\v{r}ina, Jr., A. V. Sergienko, B. M. Jost, B. E. A. Saleh, and M. C. Teich,
 Phys. Rev. A 59, 2359 (1999).
\bibitem{Joobeur1994} A. Joobeur, B. E. A. Saleh, and M. C. Teich, Phys. Rev. A. 50, 3349
(1994).
\bibitem{Joobeur1996} A. Joobeur, B. E. A. Saleh, T. S. Larchuk, and M. C. Teich, Phys. Rev. A. 53, 4360 (1996)
\bibitem{Nasr2002} M. B. Nasr, A. F. Abouraddy, M. C. Booth, B. E. A. Saleh, A. V. Sergienko, M. C. Teich,
 M. Kempe, and R. Wolleschensky, Phys. Rev. A. 65, 023816 (2002).
\bibitem{Monken1998} C. H. Monken, P. H. Souto Ribeiro, and S. Padua, Phys. Rev. A 57, 3123 (1998).
\bibitem{Walborn2004} S. P. Walborn, A. N. de Oliveira, R. S. Thebaldi, and C. H. Monken, Phys. Rev. A. 69, 023811
 (2004).
\bibitem{Molina-Terriza2005} G. Molina-Terriza, S. Minardi,
 Y. Deyanova, C.I. Osorio, M. Hendrych, and J.P. Torres,
 Phys. Rev. A {\bf 72}, 065802 (2005).
\bibitem{Shih2003} Y. Shih, Rep. Prog. Phys. 66, 1009 (2003).
\bibitem{Law2004} C. K Law and J. H. Eberly, Phys. Rev. Lett. 92, 127903 (2004).
\bibitem{Centini2005} M. Centini, J. Pe\v{r}ina Jr., L. Sciscione, C. Sibilia,
 M. Scalora, M. J. Bloemer, and M. Bertolotti, Phys. Rev. A 72, 033806 (2005).
\bibitem{PerinaJr2006} J. Pe\v{r}ina Jr., M. Centini, C. Sibilia, M. Bertolotti, and M. Scalora,
 Phys. Rev A. 73, 033823 (2006).
\bibitem{PerinaJr2008} J. Pe\v{r}ina Jr., Phys. Rev A. 77, 013803 (2008).
\bibitem{Ding1995} Y. J. Ding, S. J. Lee, and J. B. Khurgin, Phys.
 Rev. Lett. 75, 429 (1995).
\bibitem{DeRossi2002} A. De Rossi and V. Berger, Phys. Rev. Lett.
 88, 043901 (2002).
\bibitem{Booth2002} M. C. Booth, M. Atature, G. Di Giuseppe, B. E. A.
 Saleh, A. V. Sergienko, and M. C. Teich, Phys. Rev. A 66,
 023815 (2002).
\bibitem{Walton2003} Z. D. Walton, M. C. Booth, A. V. Sergienko,
 B. E. A. Saleh, and M. C. Teich, Phys. Rev. A 67, 053810 (2003).
\bibitem{Walton2004} Z. D. Walton, A. V. Sergienko, B. E. A. Saleh, and M. C. Teich,
 Phys. Rev. A 70, 052317 (2004).
\bibitem{PerinaJr2009} J. Pe\v{r}ina Jr., A. Luk\v{s}, O. Haderka, and M. Scalora, Phys. Rev. Lett. 103, 063902
 (2009).
\bibitem{PerinaJr2009a} J. Pe\v{r}ina Jr., A. Luk\v{s}, and O. Haderka, Phys. Rev. A 80, 043837
 (2009).
\bibitem{Saleh1998} B. E. A. Saleh, A. Joobeur, and M. C. Teich, Phys. Rev. A. 57, 3991 (1998).
\bibitem{Howell2004} J. C. Howell, R. S. Bennink, S. J. Bentley, and R. W. Boyd, Phys. Rev. Lett. 92, 210403 (2004).
\bibitem{DAngelo2004} M. D'Angelo, Y.-H. Kim, S. P. Kulik, and Y. Shih, Phys. Rev. Lett. 92, 233601 (2004).
\bibitem{Brambilla2004} E. Brambilla, A. Gatti, M. Bache, and L. A. Lugiato, Phys. Rev. A. 69, 023802 (2004).
\bibitem{Jost1998} B. M. Jost, A. V. Sergienko, A. F. Abouraddy, B. E.
 A. Saleh, and M. C. Teich, Opt Expr. 3, 81 (1998).
\bibitem{Haderka2005} O. Haderka, J. Pe\v{r}ina Jr., and M. Hamar, J. Opt.
 B: Quantum Semiclass. Opt. 7, S572 (2005).
\bibitem{Haderka2005a} O. Haderka, J. Pe\v{r}ina Jr., M. Hamar, and J.
 Pe\v{r}ina, Phys. Rev. A 71, 033815 (2005).
\bibitem{Jiang2003} Y. Jiang, O. Jedrkiewicz, S. Minardi, P. Di Trapani, A.
 Mosset, E. Lantz, and F. Devaux, Eur. Phys. J. D 22, 521 (2003).
\bibitem{Jedrkiewicz2004} O. Jedrkiewicz, Y.-K. Jiang, E. Brambilla, A. Gatti, M. Bache,
 L. A. Lugiato, and P. Di Trapani, Phys. Rev. Lett. 93,
 243601 (2004).
\bibitem{Jedrkiewicz2006} O. Jedrkiewicz, E. Brambilla, M. Bache, A.
 Gatti, L. A. Lugiato, and P. di Trapani, J. Mod. Opt. 53, 575
 (2006).
\bibitem{Saleh1991} B. E. A. Saleh and M. C. Teich, {\it Fundamentals of
 photonics} (Wiley, New York, 1991).
\bibitem{Ou1989} Z. Y. Ou, L. J. Wang, and L. Mandel, Phys. Rev. A 40, 1428 (1989).
\bibitem{Mandel1995} L. Mandel and E. Wolf, {\it Optical coherence and quantum optics}
 (Cambridge University press, Cambridge, 1995), Chap. 22.4.7.
\bibitem{Rubin1996} M. H. Rubin, Phys. Rev. A. 54, 5349 (1996).
\bibitem{Torres2005} J.P. Torres, F. Macia, S. Carrasco, and L.
 Torner, Opt. Lett. {\bf 30}, 314 (2005).
\bibitem{Torres2005a} J. P. Torres, M. W. Mitchell, and M. Hendrych,
 Phys. Rev. A {\bf 71}, 022320 (2005).


\end{thebibliography}
\end{document}